\documentclass[english, a4paper]{article}
\usepackage[T1]{fontenc}
\usepackage{babel}
\usepackage{amsmath}
\usepackage{amssymb}
\usepackage{graphics}
\usepackage{epsfig}

\makeatletter
\@addtoreset{chapter}{part}
\makeatother

\newcommand\lsim{\mathrel{\rlap{\lower4pt\hbox{\hskip1pt$\sim$}}
    \raise1pt\hbox{$<$}}}
\newcommand\gsim{\mathrel{\rlap{\lower4pt\hbox{\hskip1pt$\sim$}}
    \raise1pt\hbox{$>$}}}

\newcommand\be{\begin{equation}}
\newcommand\bea{\begin{eqnarray} \nonumber }
\newcommand\ee{\end{equation}}
\newcommand\eea{\end{eqnarray}}

\begin{document}

\title{Black was right:\\
Price is within a factor 2 of Value}

\author{J. P. Bouchaud, S. Ciliberti, Y. Lemp\'eri\`ere, \\
A. Majewski, P. Seager \& K. Sin Ronia \\
Capital Fund Management, \\ 
23 rue de l'Universit\'e, 75007 Paris, France \\
{\tt{jean-philippe.bouchaud@cfm.fr.}}
 }

\date{\today}
\maketitle

\begin{abstract}
We provide further evidence that markets trend on the medium term (months) and mean-revert on the long term (several years). Our results bolster Black's intuition that prices tend to be off roughly by a factor of 2, and take years to equilibrate. The story behind these results fits well with the existence of two types of behaviour in financial markets: ``chartists'', who act as trend followers, and ``fundamentalists'', who set in when the price is clearly out of line. Mean-reversion is a self-correcting mechanism, tempering (albeit only weakly) the exuberance of financial markets.

\end{abstract}

\section{Introduction}

In his remarkably insightful 1986 piece called ``Noise'', Fisher Black famously wrote \cite{Black}: {\it An efficient market is one in which price is 
within a factor 2 of value, i.e. the price is more than half of value and less than twice value.} He went on saying:  
{\it The factor of 2 is arbitrary, of course. Intuitively, though, it seems reasonable to me, in the light of sources of uncertainty about value and the strength
of the forces tending to cause price to return to value. By this definition, I think almost all markets are efficient almost all of the time.}

As far as we are concerned, we always believed that Black was essentially right, precisely for the argument he sketched: humans are pretty much clueless about the
``fundamental'' value of anything traded on markets, except perhaps in relative terms.\footnote{As noted by O. Wilde, people know the price of everything and the value of nothing.} 
The myth that ``informed'' traders step in and arbitrage away any small discrepancies between 
value and prices does not make much sense. The wisdom of crowds is too easily distracted by trends and panic \cite{McKay,Thistime,deLong}. In Black's view (see also \cite{Summers,Poterba}), prices evolve pretty 
much unbridled in response to uninformed supply and demand flows, until the difference with value is strong enough for some mean-reversion forces to  drive  prices back to more reasonable levels. 
If Black's uncertainty band $\Delta$ was -- say -- $0.1 \%$, the efficient market theory (EMT) would be a very accurate representation of reality for most purposes. 
But if $\Delta=50 \%$ or so, as Black imagined, EMT would only make sense on time scale longer than the mean-reversion time $T_{\text{MR}}$, the order of magnitude of which is set by $\sigma \sqrt{T_{\text{MR}}} \sim \Delta$. 
For stock indices with $\sigma \sim 20 \%$/year, one finds $T_{\text{MR}} \sim 6$ years. 

\begin{figure}[!htb]
\centering
\includegraphics[width=10cm]{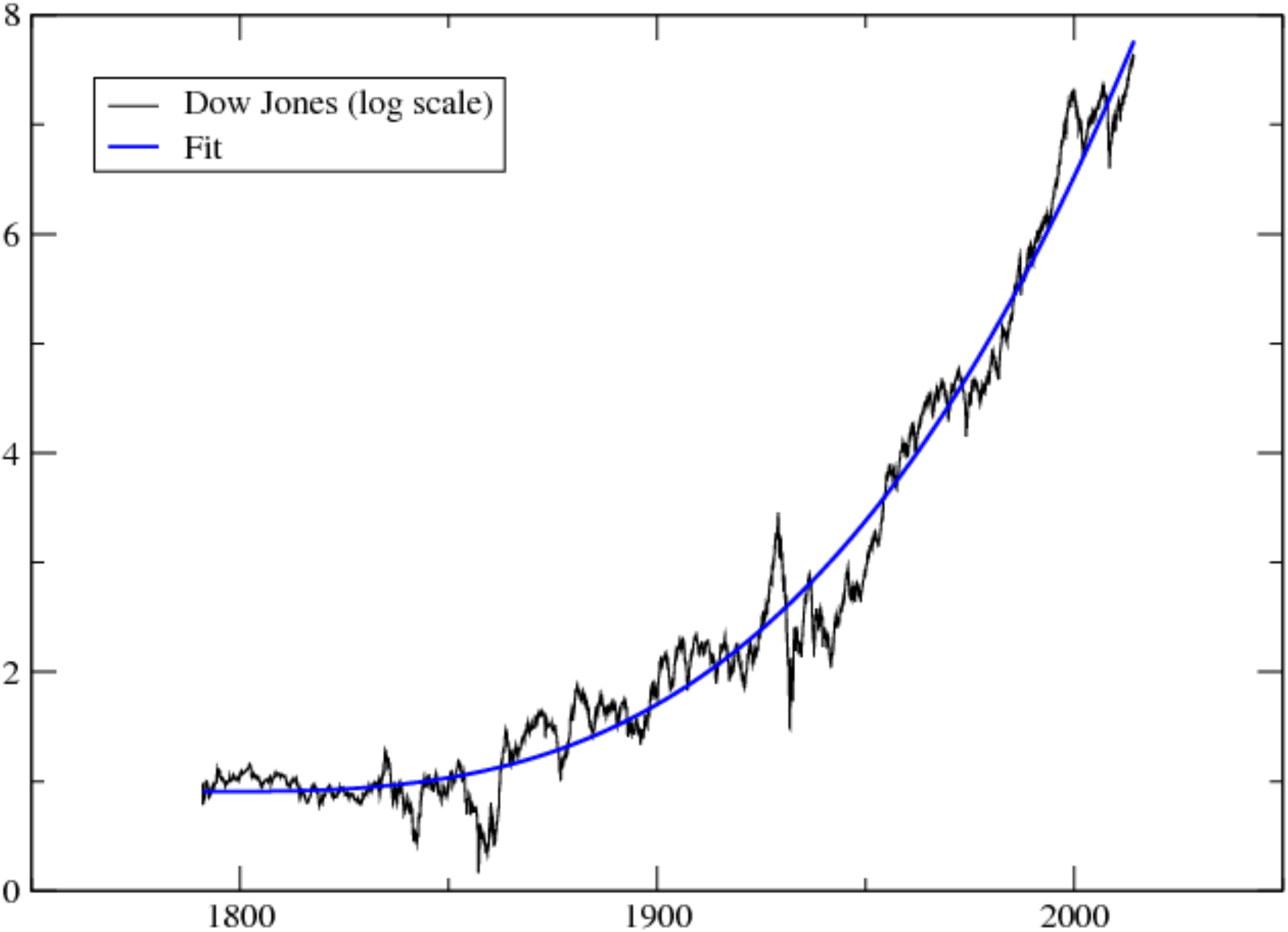}
\caption{Dow Jones Index (in log scale) since 1791, together with a non-linear long term trend, here a simple cubic fit $\propto (t-t_0)^3$ with $t_0=1791$. This suggests that the return of the stock market actually increases with time.}
\label{fig:trend}
\end{figure}

The dynamics of prices within Black's uncertainty band is in fact not random but exhibits trends: in the absence of strong fundamental anchoring forces, investors tend to under-react to news and/or take cues 
from past price changes themselves \cite{Thaler,deLong,Stein,Hirshleifer,Thesmar}. This induces positive autocorrelation of returns that have been documented in virtually all financial markets -- see e.g. \cite{Covel,Trends} and refs. therein. 
The picture that emerges, and that we test in the present study, is therefore the following: market returns are positively correlated on time scales $\ll T_{\text{MR}}$ and negatively correlated on long time 
scales $\sim T_{\text{MR}}$, before eventually following the (very) long term fate of fundamental value -- presumably a biased geometric random walk with a non-stationary drift (see Fig.~\ref{fig:trend}). We test this 
idea on a large set of instruments: indexes, bonds, FX and commodity futures since 1960 (using daily data) and spot prices since 1800 (using monthly data). Our results confirm, and make more precise, 
Black's intuition. We find in particular that mean-reversion forces start cancelling trend following forces after a time around 2 years, and mean-reversion appears to peak for channel widths $\Delta$ 
on the order of $50$ to $100 \%$, which corresponds to Black's ``factor 2''. 

In a way, our results are very intuitive: mean-reversion comes as a mitigating force against trend following that allows markets to become efficient on the very long run, as anticipated by many authors, see e.g. \cite{Poterba,Day,Stein,Lux}. However, even highly liquid markets only equilibrate on time scales of years -- and not seconds, as market efficient enthusiasts would claim. 

\section{Data}

We will use two sources of data, already exploited in our previous work on long term trend following \cite{Trends}: daily data on futures contracts since 1960 and monthly data on spot contracts since 1800. The detailed description
of the data can be found in \cite{Trends}. In a nutshell, our futures pool contains seven commodity contracts (crude oil, natural gas, corn, wheat, sugar, live cattle and copper), seven ten-year
bond contracts and seven stock index contracts (corresponding to Australia, Canada, Germany, Japan, Switzerland, the United Kingdom and the United States) and the six corresponding currency contracts. The spot contracts include 
the same commodities and stock indexes (with starting dates given in Tables \ref{IDXstart}, \ref{CMDstart}), bond prices since 1918 and currencies since 1973. We have actually checked that all the results shown below hold on an extended data set, where we use all contracts at our disposal. The data used in the current paper is from Global Financial Data (GFD).

\begin{table}
\begin{center}
\begin{tabular}{|c|c|}
\hline
country & start \\
\hline
USA & 1791 \\
Australia & 1875 \\
Canada & 1914 \\
Germany & 1870 \\
Switzerland & 1914 \\
Japan & 1914 \\
United Kingdom & 1693 \\
\hline
\end{tabular}
\caption{Starting date of the spot index monthly time-series for each country.}
\label{IDXstart}
\end{center}
\end{table}

\begin{table}
\begin{center}
\begin{tabular}{|c|c|}
\hline
commodity & start \\
\hline
Crude oil & 1859 \\
Natural Gas & 1986 \\
Corn & 1858 \\
Wheat & 1841 \\
Sugar & 1784 \\
Live Cattle & 1858 \\
Copper & 1800 \\
\hline
\end{tabular}
\caption{Starting date of the spot price for each commodity.}
\label{CMDstart}
\end{center}
\end{table}

\section{Predictability curves over different time horizons}

\subsection{From trends to mean-reversion}

In the following, we will study the statistical structure of the relation between past de-trended returns on scale $\tau_<$ and future de-trended returns on scale $\tau_>$. 
More precisely, define $p(t)$ as the price level of any asset (stock index, bond, commodity, etc.) at time $t$. The long term trend (seen from $t$) over some time scale $T$ is defined as:
\be
\mu_t := \frac{1}{T} \log \left[\frac{p(t)}{p(t-T)}\right].
\ee
For each contract and time $t$, we associate a point $(x,y)$ where $x$ is the de-trended past return on scale $\tau_<$ and $y$ the de-trended future return on scale $\tau_>$:
\be
x:= \log p(t) - \log p(t-\tau_<) - \mu_t \tau_<; \qquad y: =\log p(t+\tau_>) - \log p(t) - \mu_t \tau_>. 
\ee
Note that the future return is de-trended in a causal way, i.e. no future information is used here (otherwise mean-reversion would be trivial). 
For convenience, both $x$ and $y$ are normalised such that their variance is unity. We choose $T=20$ years and consider both:
\begin{itemize} 
\item Spot prices {\it monthly} data, extending back 200 years,  with $\tau_< = 5,10,15,20,25,30,40,50$ and $60$ months for backward looking horizons and choose $\tau_>=\tau_</5$.
\item Futures contracts {\it daily} data, extending back to 1960, with  $\tau_< = 10,20,40,80$, $160,320,480,640,960$ and $1280$ days for backward looking horizons and choose again $\tau_>=\tau_</5$.
\end{itemize}
In the following, we want to study how past returns on a variety of time scales from 10 days to five years predict future returns over two days to one year (respectively). The choice $\tau_>=\tau_</5$ is not critical to our conclusion, as we could have chosen $\tau_>= a \tau_<$ with $a \leq 1$ with similar results. Remarkably, futures and spot data lead to the same overall conclusions. 

\begin{figure}[!htb]
\centering
\includegraphics[width=10cm]{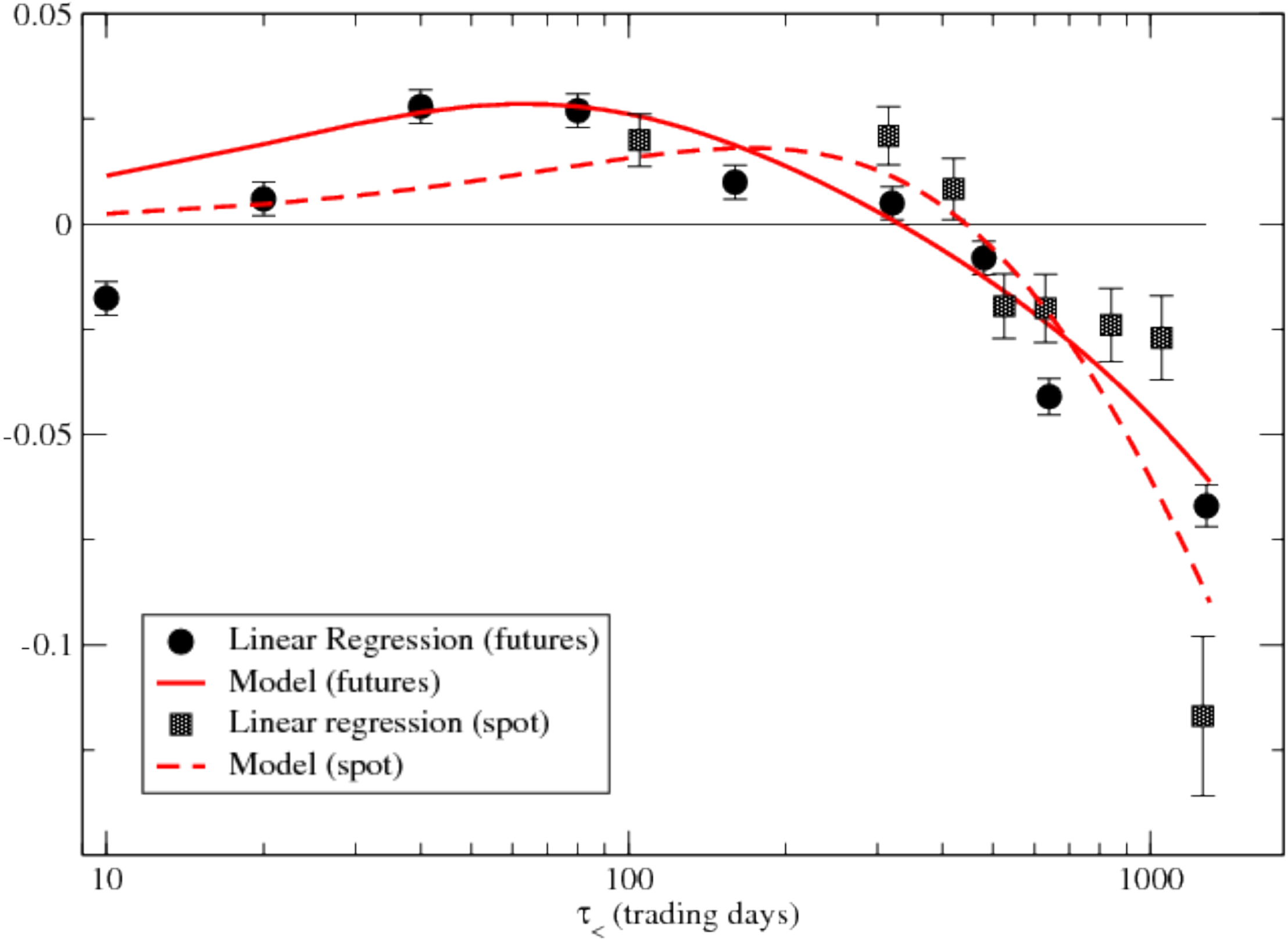}
\caption{Slope of a simple linear regression as a function of past horizon $\tau_<$ (in log scale) for (black symbols) futures daily data and (grey symbols) spot monthly data. Plain and dashed lines: Result of a simple model with medium term trends and long term mean-reversion, see Eq. \ref{theory}. Note that correlation changes sign around $\tau_< = 2$ years (500 trading days) for both data sets.}
\label{fig:fit_fut}
\end{figure}

We group all contracts in two pools: one with futures and one with spots. For each each pool we group points $(x,y)$ in sub-pools associated with different $\tau_{<}$. Then we fit linear regression in each sub-pool, removing outliers  (i.e. $|x|>4$ or $|y|>4$). The slopes of this regression are shown in Fig. \ref{fig:fit_fut}. 
For short lags $\tau_< \lsim$ 2 years, the slope is positive, compatible with the well known trend following effect ($\tau_< = 5$ months exactly recovers the results of \cite{Trends} for spot prices). But for longer lags, a clear mean-reversion effect is observed, with a negative slope increasing (in absolute value) with time horizon. The pattern is very similar for spot and futures data, although in the latter case some evidence for short term mean-reversion when $\tau_< = 10$ days can be detected. The appearance of long-term reversion on multi-year time scales was first evidenced by Poterba and Summers in the case of stock indices \cite{Poterba}. 

\subsection{A simple model}

It is insightful to compare the behaviour of the regression slope shown in Fig. \ref{fig:fit_fut} with a simple model. Assume that the de-trended log-price $\pi(t)$ evolves as a mean-reverting Ornstein-Uhlenbeck process driven by a positively correlated (trending) noise $\eta$, to wit:
\be \label{model}
\frac{d\pi(t)}{dt} = - \kappa \pi(t) + \eta(t); \qquad \langle \eta(t') \eta(t'') \rangle = 2 \sigma^2 \kappa \left[ \delta(t'-t'') + \frac{g}{2} (\gamma + \kappa) e^{-\gamma |t'-t''|} \right],
\ee
where $\kappa^{-1}$ is the mean-reversion time, $\gamma^{-1}$ the trend correlation time and $g$ a parameter measuring the strength of the trend. One can compute analytically the regression slope $s(\tau_<,\tau_>)$ of future returns on scale $\tau_>$ as a function of the past returns on scale $\tau_<$. After some manipulations (see Appendix), one finds:
\be \label{theory}
s(\tau_<,\tau_>) = \frac{C(\tau_<)+C(\tau_>)-C(\tau_<+\tau_>)-1}{2 \sqrt{(1-C(\tau_<))(1-C(\tau_>))}}
\ee
where the autocorrelation function $C$ of the process $\pi$ is defined as: 
\be
C(u):= \frac{1}{1+g} e^{-\kappa u} + \frac{g}{1+g} \frac{\gamma e^{-\kappa u} - \kappa e^{-\gamma u}}{\gamma - \kappa}.
\ee
We show as a dashed line in Fig. \ref{fig:fit_fut} the prediction of such a model with $g=0.22$, $\kappa^{-1}=16$ years and $\gamma^{-1}=33$ days, chosen to fit the futures data and $g=0.33$, $\kappa^{-1}=8$ years and $\gamma^{-1}=200$ days, chosen to fit the spot data. In view of the simplicity of the model, the qualitative agreement is quite remarkable. Although the value of the parameters $g,\kappa$ and $\gamma$ are not determined with high accuracy, their order of magnitude is reasonable. Note that by construction our stylized model is unable to account for any short-term mean-reversion, and significant discrepancies are indeed observed for $\tau_< \lsim 20$ days for which the slope $s$ becomes negative.

If the model defined by Equation \eqref{model} is taken seriously, the mean square fluctuations of the log-price around its equilibrium value is easily computed to be:
\be
\langle \pi^2 \rangle = \sigma^2 (1+g),
\ee
which we define as the square of the width of Black's uncertainty band, $\Delta^2$. But since the short term volatility of prices is simply given by $\sqrt{2\kappa} \sigma$ in this model, one can infer the value of $\sigma$ from market data. Taking a typical daily volatility of $1 \%$, one finds $\sigma^2 = 0.2$ and $\Delta \approx 0.5$ for futures data and  $\sigma^2 = 0.1$ and $\Delta \approx 0.35$  -- corresponding to prices erring by a factor $\sim 1.5$ from their ``reference'' value, much as Black argued \cite{Black}. While the precise value of $\Delta$ should not be taken too seriously, we find our results quite suggestive.

\begin{figure}[!htb]
\centering
\includegraphics[width=10cm]{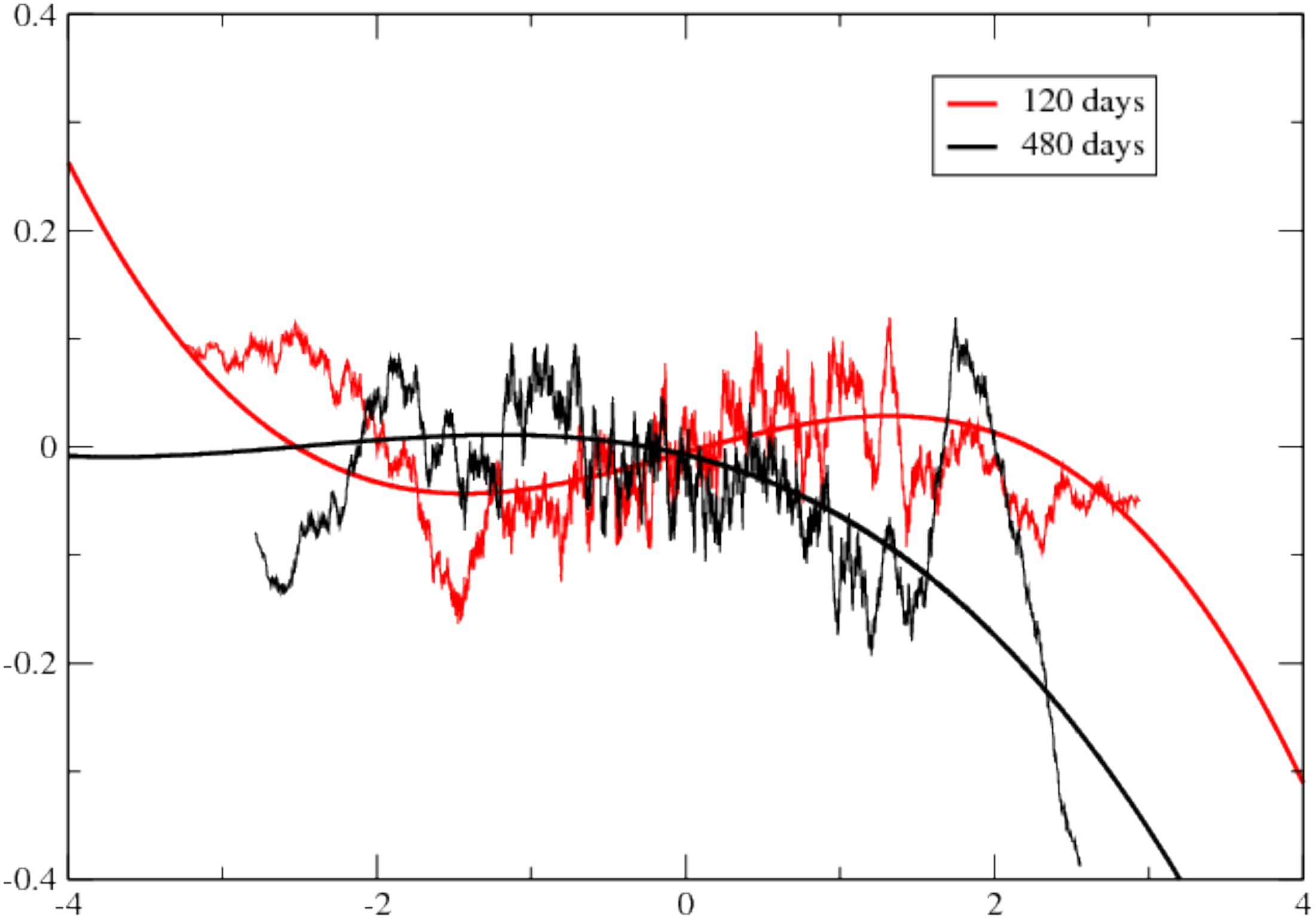}
\caption{Plot of $y$ (de-trended future return on scale $\tau_>=\tau_</5$) vs. $x$ (de-trended past return on scale $\tau_<$) for futures daily data, and $\tau_<=120$ business days (red lines) and $\tau_<=480$ business days (black lines). We show both a running average through 200 consecutive points, and a cubic fit of the raw data (excluding points beyond 4-$\sigma$). The change of slope sign as $\tau_<$ increases, and the presence of non-linear effects, are clearly suggested by these plots.}
\label{fig:cubics}
\end{figure}

\subsection{Non-linear effects?}

\begin{figure}[!htb]
\centering
\includegraphics[width=6cm]{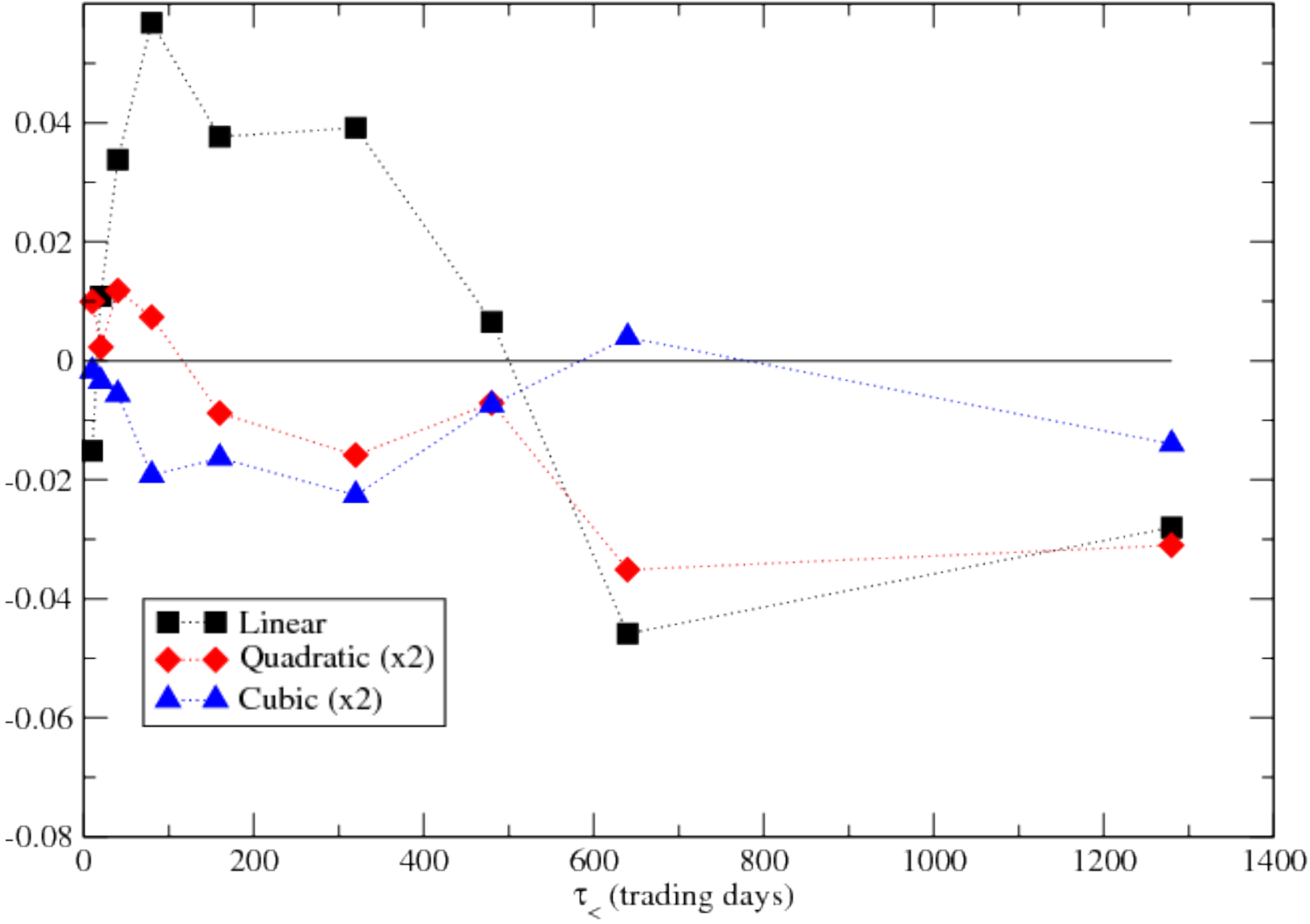} 
\includegraphics[width=6cm]{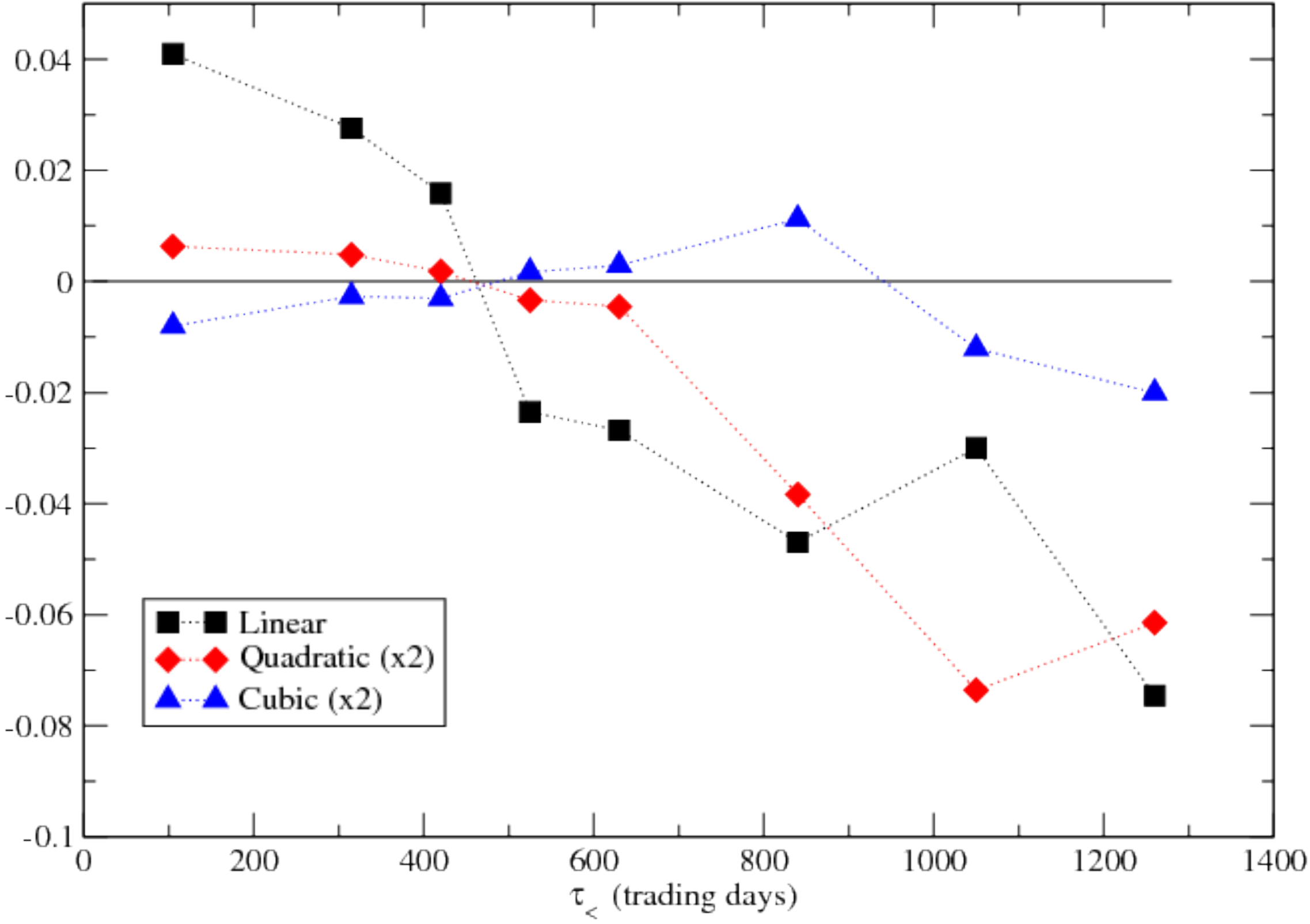}
\caption{Cubic fit parameters as a function of past horizon $\tau_<$ for (left) futures daily data and (right) spot monthly data. Note that the linear coefficient changes sign around $\tau_< = 2$ years (500 trading days), as is found for the slope of the simple linear regression (see Fig. \ref{fig:fit_fut}).}
\label{fig:fit_cubic}
\end{figure}

A closer look at the  plot $(x,y)$ however reveals significant departure from a simple linear behaviour -- see Fig. \ref{fig:cubics}. One expects trend effects to weaken as the absolute value of past returns increases, as indeed reported in \cite{Trends}. We have therefore attempted a cubic polynomial regression, devised to capture both potential asymmetries between positive and negative returns, and saturation or even inversion effects for large returns. The linear, quadratic and cubic coefficients of the fit are also shown in Fig. \ref{fig:fit_cubic}. The linear coefficient of the fit behaves very similarly to the slope of the simple linear regression, as expected. Our conclusion on the change of sign of the slope around $\tau_< = 2$ years is therefore robust. The quadratic term, on the other hand, is positive for short lags but becomes negative at longer lags, for both data sets. The cubic term appears to be negative for all time scales in the case of futures, but this conclusion is less clear-cut for spot data. 

The behaviour of the quadratic term is interesting, as it indicates that positive trends are stronger than negative trends on short time scales, while negative trends are stronger than positive trends on long time scales. A negative cubic term, on the other hand, suggests that large moves (in absolute value) tend to mean-revert, as expected, even on short time scales where trend is dominant for small moves. Taking these non-linearities into account however does not affect much the time scale for which the linear coefficient vanishes, i.e. roughly two years. 

\section{Conclusion}

In this paper, we have provided some further evidence that markets trend on the medium term (months) and mean-revert on the long term (several years). This dovetails with Black's intuition that prices tend to be off by a factor of two: it takes roughly six years for the price of an asset with 20 \% annual volatility to vary by 50 \%. 

Such a long time scale is another nail in the coffin of efficient market theory: as anticipated by Summers and others \cite{Summers,Poterba}, it is not because returns are only weakly correlated on short time scales that prices efficiently reflect value. The story behind our results, on the other hand, fits well with a series of previous models \cite{Day,Lux,GB,Hommes,Chiarella}, which postulate the presence of two types of agents in financial markets: ``chartists'', who act as trend followers, and ``fundamentalists'', who set in when the price is clearly out of whack. Mean-reversion is a self-correcting mechanism, tempering (albeit only weakly) the exuberance of financial markets. 

From a very practical point of view, our results suggest that universal trend following strategies should be supplemented by universal price-based ``value'' strategies that mean-revert on long term returns.\footnote{Value strategies based on more fundamental (but less universal) indicators have been discussed in \cite{AQR}.} As is well known, trend following strategies offer an hedge against market drawdowns (see e.g. \cite{Lam}); value strategies offer a hedge against over-exploited trends. As a consequence, we find that mixing both strategies significantly improves the profitability of the resulting portfolios.  

\vskip 1cm

Acknowledgments: We thank L. Duchayne, A. Rej and M. Potters for useful insights, and A. Breedt for comments. 

\section*{Appendix}

The dynamics of process $\pi$ is given by Eq. (\ref{model}). Using variation of parameters we can write
\begin{equation}\label{eq:VariationOfParameters}
d \left( \mathrm{e}^{\kappa t} \pi (t) \right) =  \kappa\mathrm{e}^{\kappa t} \pi (t) dt+ \mathrm{e}^{\kappa t} d \pi (t) = \mathrm{e}^{\kappa t} d \eta (t).
\end{equation}
From (\ref{eq:VariationOfParameters}) we get 
\begin{equation}
\pi(t) = \mathrm{e}^{-\kappa t} \pi_0  + \mathrm{e}^{-\kappa t} \int_0^t \mathrm{e}^{\kappa s} d \eta(s).
\end{equation}
For large $t$,  $\pi(t)$ admits stationary distribution with zero mean. Lets compute covariance of the process:
\begin{equation}
\begin{split}
\langle \pi(t') \pi(t'') \rangle &= \pi_0 \mathrm{e}^{-\kappa (t' + t'')} + \mathrm{e}^{-\kappa (t' + t'')} \int_0^{t'} \int_0^{t''}\mathrm{e}^{\kappa (s' + s'')} \langle \eta(s') \eta (s'') \rangle ds'' ds'\\
&= \pi_0 \mathrm{e}^{-\kappa (t' + t'')} + 2  \kappa\sigma^2 \mathrm{e}^{-\kappa (t' + t'')} \int_0^{t'} \int_0^{t''}\mathrm{e}^{\kappa (s' + s'')} \delta(s'-s'')ds''ds'\\
&+  \kappa g (\gamma + \kappa)\sigma^2 \mathrm{e}^{-\kappa (t' + t'')} \int_0^{t'} \int_0^{t''}\mathrm{e}^{\kappa (s' + s'')} \mathrm{e}^{-\gamma |s' - s''|}ds''ds'\\
&= \pi_0 \mathrm{e}^{-\kappa (t' + t'')} + \sigma^2 \left( \mathrm{e}^{-\kappa |t' - t''|}  -  \mathrm{e}^{-\kappa (t' + t'')} \right)\\
&+\frac{\kappa g \sigma^2}{\kappa - \gamma}  \left( \mathrm{e}^{-\gamma t'' - \kappa t'}
+ \mathrm{e}^{-\gamma t' - \kappa t''}\right)+\frac{ g \sigma^2}{\kappa - \gamma} \left( \kappa \mathrm{e}^{-\gamma |t'-t''|} - \gamma \mathrm{e}^{-\kappa |t' - t''|} \right).
\end{split}
\end{equation}

For large $t', t''$ we thus obtain that 

\begin{equation}\label{eq:covariance}
\langle \pi(t') \pi(t'') \rangle =  \sigma^2 \left( \mathrm{e}^{-\kappa |t' - t''|}   + \frac{ g }{\gamma-\kappa} \left(  \gamma \mathrm{e}^{-\kappa |t' - t''|} - \kappa \mathrm{e}^{-\gamma |t'-t''|}  \right) \right)
\end{equation}
and the variance of the process is equal $\sigma^2 (1+g)$. Then, it is straightforward to obtain the slope of 
linear regression of $\pi(t+\tau_{>})-\pi(t)$ on $\pi(t)-\pi(t-\tau_{<})$ from the covariance of $\pi(t)$, with the result given in the main text.

\end{document}